\documentclass{ifacconf}

\usepackage{amssymb}
\usepackage{amsmath}
\usepackage{subcaption}
\usepackage{xcolor}
\usepackage{url}

\newcommand{\R}{\mathbb{R}}

\newcommand{\norm}[1]{\left\lVert#1\right\rVert}

\usepackage{graphicx}      
\usepackage{natbib}        
\begin{document}
\begin{frontmatter}
\thanks{This research work is supported in part by NSERC-DG RGPIN-2020-04759 and Fonds de recherche du Qu\'ebec (FRQ).}
\title{Perception-Limited Smooth Safety Filtering} 

\author[Quebec]{Lyes Smaili}
\author[Quebec,Ontario]{Soulaimane Berkane} 

\address[Quebec]{Department of Computer Science and Engineering, University of Quebec in Outaouais, Gatineau, QC, Canada. (e-mail: {smal01, Soulaimane.Berkane}@uqo.ca).}
\address[Ontario]{Department of Electrical Engineering, Lakehead University, ON, Canada.}

\begin{abstract}
This paper develops a smooth safety-filtering framework for nonlinear control-affine systems under limited perception. Classical Control Barrier Function (CBF) filters assume global availability of the safety function—its value and gradient must be known everywhere—an assumption incompatible with sensing-limited settings, and the resulting filters often exhibit nonsmooth switching when constraints activate. We propose two complementary perception-aware safety filters applicable to general control-invariant safety sets. The first introduces a smooth perception gate that modulates barrier constraints based on sensing range, yielding a closed-form Lipschitz-safe controller with forward-invariance guarantees. The second replaces the hard CBF constraint with a differentiable penalty term, leading to a smooth unconstrained optimization-based safety filter consistent with CBF principles. For both designs, we establish existence, uniqueness, and forward invariance of the closed-loop trajectories. Numerical results demonstrate that the proposed smooth filters enable the synthesis of higher-order tracking controllers for systems such as drones and second-order ground robots, offering substantially smoother and more robust safety-critical behaviors than classical CBF-based filters.
\end{abstract}


\end{frontmatter}
\section{Introduction} 
\subsection{Motivation}
Safety is a fundamental requirement in control of autonomous systems operating in dynamic and uncertain environments. Control Barrier Functions (CBFs) have emerged as a powerful tool for synthesizing minimally invasive safety filters that render a desired set forward invariant while preserving a nominal control input whenever safety permits. In their standard form, CBF filters are implemented as solutions to strictly convex quadratic programs (QPs) that minimally perturb the nominal control subject to an affine constraint on the instantaneous barrier derivative. However, classical CBF implementations implicitly assume \emph{full perception} of the environment, and switch between the nominal control and a constrained projection when the barrier constraint becomes active. This can be overly conservative in perception-limited settings, where the controller must operate with incomplete knowledge of the environment, and can lead to undesirable nonsmoothness. This work overcomes these limitations by introducing smooth, perception-aware safety filters that relax the full-perception assumption and eliminate the switching behavior of classical CBF-QP controllers.

\subsection{Related Work}
Barrier certificates were first introduced by \citet{WIELAND2007462} and later developed into a unified CBF-CLF framework by \citet{Ames2017}, with a comprehensive overview given in \citet{Ames2019}. CBFs have since been applied in a range of settings, including multi-robot collision avoidance \citep{Wang2017} and safe stabilization via CLF-CBF combinations \citep{ROMDLONY201639}.

QP-based CBF filters \citep{Ames2017} have become the standard mechanism for real-time safety filtering, as they minimally modify a nominal controller while ensuring forward invariance. However, recent analyses have shown that these filters may introduce \emph{undesired equilibria}, sometimes asymptotically stable, which can compromise goal achievement \citep{Reis2021,TAN2024111359,Mestres2025}. Subsequent studies characterized the existence, multiplicity, and stability of such equilibria \citep{Chen2024,Cortez2022} and examined continuity and boundedness of minimum-norm CBF controllers \citep{Alyaseen2025}, underscoring the importance of regularity and well-posedness.

A parallel line of work addresses robustness and uncertainty. Robust CBF formulations \citep{JANKOVIC2018,Kolathaya2019} and GP-based safety filters \citep{Castaneda2021,MESTRES2024111800} enhance feasibility under uncertain dynamics, while reduced-order, model-free, and inverse-optimality approaches \citep{Molnar2022MFSCCR,Molnar2023SCCBIROM,Krstic2024} broaden the applicability of safety-critical control. However, these approaches typically preserve the nonsmooth switching behavior inherent to QP-based CBF implementations.

More recently, the need for \emph{smooth} safety filters has gained increasing attention. Classical CBF-QP controllers are piecewise smooth and switch sharply as constraints activate, producing control discontinuities that complicate tracking tasks, degrade performance for high-order systems (e.g., drones and second-order robots), and become particularly problematic under perception limitations, where constraint activation must occur progressively as obstacles enter the sensing region. To mitigate these issues, smooth universal formulas such as Sontag’s construction \citep{SONTAG1989} have been adapted to safety filtering \citep{Li2023GIUFSS,Kolathaya2019,Krstic2024,MESTRES2024111800}, and \citet{Cohen2023CSSSFIFT} employed the Implicit Function Theorem to characterize smooth CBF-based controllers. Other approaches include weighted-centroid control laws \citep{Ong2019}, barrier backstepping \citep{Taylor2022}, model-free safety-critical controllers \citep{Molnar2022MFSCCR}, and penalty-based safety filters \citep{Mestres2023}. These methods improve smoothness but do not directly address the coupled need for smooth activation and perception-limited safety enforcement, which is central to the present work.
\subsection{Contributions and Organization}
This paper develops a unifying smooth and perception-aware framework for safety filtering in nonlinear control-affine systems. First, we introduce a perception-limited CBF filter that employs a smooth gating mechanism to activate barrier constraints only within the sensing range; a closed-form controller is obtained, and its Lipschitz continuity and forward-invariance properties are formally established. Second, we propose a smooth penalty-based formulation that substitutes the hard inequality constraint with a differentiable penalty term, yielding a fully smooth closed-form controller that maintains safety while providing higher regularity. Finally, we present a unified theoretical analysis establishing existence, uniqueness, and forward invariance of the closed-loop trajectories under both filters, together with asymptotic stability of the safety set.

The remainder of the paper is organized as follows. 
Section~\ref{sec:preliminaries} reviews the necessary preliminaries on control-affine systems and introduces the perception-limited CBF filter and its theoretical properties. Section~\ref{sec:penalty-based} introduces the smooth penalty-based formulation and establishes its smoothness and safety guarantees. Section~\ref{sec:simulation} provides illustrative examples and numerical results. 
Concluding remarks are given in Section~\ref{sec:conclusion}.

\paragraph*{Notation.}
Let $\mathbb{N}$ and $\R$ denote the set of natural and real numbers, respectively. Let $\R^n$ be the $n$-dimensional Euclidean space, with $n\in\mathbb N$. For a given vector $x\in\R^n$, we denote by $\|x\|$ and $\|x\|_A$ the Euclidean norm and weighted norm operators, respectively, where $A\in\R^{n\times n}$. A function $f$ is said to be of class $\mathcal{C}^k$ if its derivatives $f',f'',\cdots,f^k$ exist and are continuous. Let $\mathcal{B}(x,r):=\{y\in\R^n\,|\,\|x-y\| < r\}$ denote the Euclidean ball of radius $r>0$ and centered at $x\in\R^n$. A continuous function $f:[0,a]\to[0,+\infty)$ is a class $\mathcal{K}$ function if it is strictly increasing and $f(0)=0$, while it is a class $\mathcal{K}_\infty$ if $a\to+\infty$ and $\lim_{x\to+\infty} f(x) = +\infty$. Let $\mathcal{C}(f)$ denote the class of the function $f$. The notation $L_gf(x)$ denotes the Lie  derivative of $f:\R^n\to\R$ along the vector field $g:\R^n\to\R^{n\times m}$ with respect to $x$, \emph{i.e.,} $L_gf(x) = \frac{\partial f}{\partial x}g(x)$.

\section{Perception-Limited Safety Filtering}\label{sec:preliminaries}
In this section, we formalize the notion of safety filtering for control-affine systems and introduce a perception-limited extension of the classical CBF condition. Although the results are largely intuitive, they serve as a benchmark formulation for safety filtering under limited perception and motivate the smooth penalty-based construction developed in the next section.

We begin by recalling the standard barrier-based formulation for nonlinear control-affine systems of the form
\begin{equation}
\dot{x} = f(x) + g(x) u, 
\quad x \in \R^n, \quad u \in \R^m,
\label{eq:sys}
\end{equation}
where $f$ and $g$ are locally Lipschitz. We consider a nominal feedback law $u_0:\mathbb{R}^n \to \mathbb{R}^m$ of class $\mathcal{C}^k$ with $k \ge 1$, representing the desired control input in the absence of safety constraints.
We encode safety through a function $h:\mathbb{R}^n \to \mathbb{R}$ of class $\mathcal{C}^l$ with $l \ge 1$, and define the \emph{safe set}
\begin{equation}\label{eq:safe set}
\mathcal{X} := \{\,x \in \mathbb{R}^n \mid h(x) \ge 0\,\},
\end{equation}
which we seek to render forward invariant under a suitably modified control law. In the classical CBF framework, forward invariance of $\mathcal{X}$ is ensured by imposing
\begin{equation}
\dot{h}(x,u) + \alpha\!\left(h(x)\right) \ge 0,
\label{eq:cbf-cond}
\end{equation}
for some extended class-$\mathcal{K}$ function $\alpha : [0,\infty) \to \mathbb{R}$ of class $\mathcal{C}^p$, which provides an increasing safety margin as $h$ approaches zero.
 The derivative of $h$ along the trajectories of \eqref{eq:sys} is
\begin{equation}
\dot{h}(x,u) = L_f h(x) + L_g h(x)\,u 
= c(x) + a(x) u,
\label{eq:hdot}
\end{equation}
where $c(x) := L_f h(x)$ and $ a(x) := L_g h(x) \in \mathbb{R}^{1\times m}$.
To guarantee that the constraint \eqref{eq:cbf-cond} is enforceable through the control input, we adopt the standard assumption that $h$ has relative degree one with respect to the system dynamics.
\begin{assum}
\label{assumption:RelativeDegreeOne}
The barrier function $h$ has relative degree one with respect to system \eqref{eq:sys}, \textit{i.e., }
\[
L_g h(x) = a(x) \neq 0, \quad \forall x \in \mathcal{X}.
\]
\end{assum}

We propose a perception-limited modification on the CBF condition \eqref{eq:cbf-cond} by introducing a smooth gate that activates the barrier inequality only within a sensing band, \emph{i.e.,} $h(x)\le\delta$.

\begin{defn}[Smooth perception gate]
    Let $\delta>\epsilon>0$ be two real numbers. A smooth and non-increasing function $\gamma:[0,+\infty)\to[0,1]$ is a perception gate if satisfies the following properties:
    \begin{enumerate}
        \item $\gamma$ is a $\mathcal{C}^q$-class function on $[0,+\infty)$;
        \item $\gamma(h)=0$ for all $h\ge\delta$;
        \item $\gamma(h)=1$ for all $h\le\epsilon$.
    \end{enumerate}
\end{defn}
This gating function allows the safety constraint to be fully active when the system is close to to the boundary $\partial\mathcal{X}$, while smoothly deactivating it as the state moves beyond the perception threshold. With this, we can write the \emph{gated barrier inequality} as follows:
\begin{equation}
\gamma\!\big(h(x)\big)\,\Big(c(x)+a(x)u+\alpha\!\big(h(x)\big)\Big)\ \ge\ 0.
\label{eq:gated-cbf-ineq}
\end{equation}
Consequently, the resulting \emph{perception-limited} CBF filter preserves nominal control whenever the system is far from constraints, reducing unnecessary conservatism. We formalize the gated safety filter as the solution of the following quadratic program:
\begin{equation}
\label{eq:gated-cbf-qp}
\begin{aligned}
u^\star(x)\in\arg&\min_{u\in\R^m}\quad  \tfrac12 \norm{u-u_0(x)}_{W}^2\\
\text{s.t.}\quad & \gamma(h(x))\big(c(x)+a(x)u+\alpha(h(x))\big)\ge 0,
\end{aligned}
\end{equation}
where $W\succ 0$ is a weighting matrix.
\begin{lem}\label{lemma:Gated-QP-Solution}
    The solution of the gated QP problem \eqref{eq:gated-cbf-qp} is given by:
    \begin{equation}
    u^\star(x) =
    \begin{cases}
    u_0(x), & \text{if } \gamma(h(x))\sigma(x)\ge 0, \\[1pt]
    u_0(x) - \dfrac{\sigma(x)}{\|a(x)\|_{W^{-1}}}\,\nu(x), & \text{if } \gamma(h(x))\sigma(x)<0,
    \end{cases}
    \label{eq:gated-qp-solution}
    \end{equation}
    where
    \begin{align}
        \sigma(x) &:= c(x)+a(x)u_0(x)+\alpha(h(x)),\label{eq:sigma_1}\\
        \nu(x)&:=W^{-1}a(x)^\top.\label{eq:v x}
    \end{align}
\end{lem}
\begin{pf}
    The problem is a strictly convex quadratic program with one linear inequality constraint. Its solution is obtained directly from the KKT optimality conditions, as detailed in classical convex optimization references such as \cite{boyd2004convex}.
\end{pf}

Lemma~\ref{lemma:Gated-QP-Solution} shows that the perception-limited safety filter retains the nominal control action $u_0(x)$ whenever the gated barrier constraint is inactive, \emph{i.e.,} when the system is either far from the safe set boundary or the nominal input already satisfies the inequality. When the constraint becomes active, the filter minimally adjusts the control input. This closed-form solution exploits the fact that for $h(x) \gg 0$ the system is far from any perceived constraint. For instance, we denote by $\mathcal{X}_\delta$ the \emph{perception-free} region, given by
\begin{equation}
    \mathcal{X}_\delta := \{\,x\in\R^n \mid h(x) \ge \delta\,\}.
\end{equation}
where $\delta$ is representing the range beyond which safety constraints are deactivated.

It is worth mentioning that the classical CBF formulation and its solution can be easily recovered by taking $\gamma(h(x))=1$ for all $x\in\mathcal{X}$.

The following theorem addresses the regularity and uniqueness of the closed-form solution \eqref{eq:gated-qp-solution}.
\begin{thm}\label{thm:gated-invariance}
Consider the safe set $\mathcal{X}$ defined by \eqref{eq:safe set}. Then the feedback control law \eqref{eq:gated-qp-solution} is locally Lipschitz-continuous on $\mathcal{X}$ under which the dynamical system \eqref{eq:sys} admits a unique solution for all initial condition $x(0)\in\mathcal{X}$ and $\mathcal{X}$ is forward invariant.
\end{thm}
\begin{pf}
    See Appendix \ref{appendix:B}.
\end{pf}

Theorem~\ref{thm:gated-invariance} establishes two fundamental properties of the perception-limited CBF filter. First, the closed-form control law \eqref{eq:gated-qp-solution} is locally Lipschitz, as it results from a piecewise definition that switches between two cases depending on the active inequality. Second, forward invariance of the safe set $\mathcal{X}$ is guaranteed despite the presence of the perception gate, which selectively deactivates safety constraints when obstacles lie outside the sensing range. Thus, incorporating smooth gating does not compromise existence and uniqueness of solutions, and safety guarantees are preserved.

Nevertheless, the perception-limited QP formulation remains a \emph{hard-constrained} optimization problem, and the resulting control law is obtained through discontinuous case distinctions when the constraint becomes active. Although Lipschitz continuity is preserved, the piecewise structure may still limit smoothness, particularly in tracking or higher-order systems where differentiability of the control input is essential. These considerations motivate the development of a fully smooth alternative. The next section introduces a penalty-based filter that replaces the hard constraint with a differentiable penalty term, yielding a smooth closed-form safety filter.

\begin{figure}[t]
    \centering
    \begin{subfigure}[t]{0.45\linewidth}
        \centering
        \includegraphics[width=\linewidth]{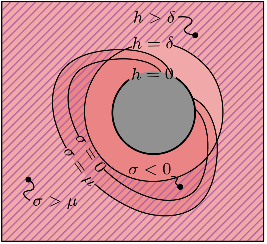}
        \caption{}
        \label{subfig:a}
    \end{subfigure}%
    ~ 
    \begin{subfigure}[t]{0.45\linewidth}
        \centering
        \includegraphics[width=\linewidth]{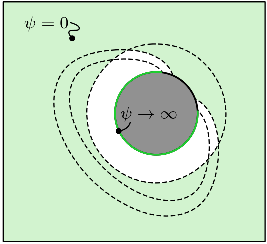}
        \caption{}
        \label{subfig:b}
    \end{subfigure}
    ~
    \begin{subfigure}[t]{\linewidth}
        \centering
        \includegraphics[trim={0.7cm 0.10cm 0.7cm 1.1cm},clip,width=0.8\linewidth]{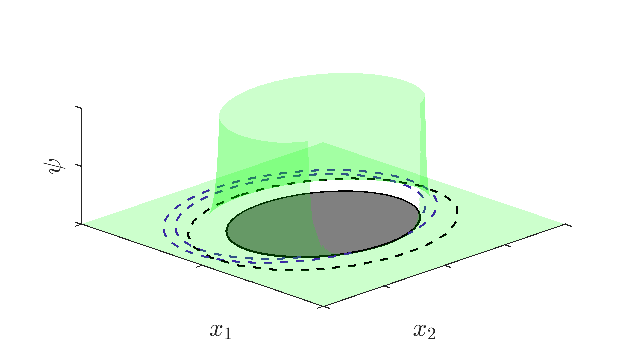}
        \caption{}
        \label{subfig:c}
    \end{subfigure}
    \caption{Visualization of the smooth penalty function $\psi(h,\sigma)$. (\subref{subfig:a}) shows the level sets of the functions $h$ and $\sigma$, where the hashed region represents the region for $h>\delta$. (\subref{subfig:b}) is the corresponding values of the function $\psi(h,\sigma)$. As provided in Definition \ref{definition:SmoothPenaltyFunction}, $\psi(h,\sigma) = 0$ whenever $h \ge \delta$ or $\sigma \ge \mu$, and grows unbounded as $h \to 0$ with $\sigma \le 0$. In all other regions, $\psi$ takes smooth values in $(0, +\infty)$. (\subref{subfig:c}) is a 3D visualization of $\psi(h,\sigma)$.}
    \label{fig:psi}
\end{figure}
\section{Smooth Penalty-Based Filter with Perception Limitation}\label{sec:penalty-based}
The perception-limited CBF filter introduced earlier ensures forward invariance through a hard inequality constraint, but the resulting piecewise structure limits the achievable smoothness of the control law. To overcome this limitation, and building on the smooth safety-filtering approach recently proposed for single-integrator systems in \cite{smaili2025smooth}, we introduce an alternative formulation in which the gated safety constraint in \eqref{eq:gated-cbf-qp} is replaced by a \emph{smooth, unconstrained} optimization problem. Central to this construction is the introduction of a \emph{penalty function} defined as follows:
\begin{defn}[Smooth Penalty Function]\label{definition:SmoothPenaltyFunction}
Let $\mu>0$ and $\delta>0$ be given constants. A function $\psi:\mathbb{R}\times\mathbb{R}\to\mathbb{R}_{\ge 0}$ is called a smooth penalty function if:
\begin{enumerate}
    \item $\psi$ is of class $\mathcal{C}^r$ on $\mathbb{R}\times\mathbb{R}$;
    \item $\psi(h,\sigma)=0$ whenever $h\ge \delta$ or $\sigma\ge \mu$;
    \item $\psi(h,\sigma)\to+\infty$ as $h<0$ and $\sigma<0$ simultaneously.
\end{enumerate}
\end{defn}

We use this penalty function to penalize the nominal controller $u_0(x)$ whenever it violates the gated safety condition (See Figure~\ref{fig:psi}). Accordingly, we consider the unconstrained optimization problem
\begin{equation}
\label{eq:penalty-problem}
u^\star(x)\in\arg\min_{u\in\mathbb{R}^m}
\;\frac12\|u-u_0(x)\|_{W}^2
\;+\;\frac12\,P(x,u),
\end{equation}
where the penalty term is defined as
\begin{equation}
P(x,u):=\psi\!\left(h(x),\sigma(x,u)\right)\,\big[\sigma(x,u)\big]^2,
\end{equation}
with
\begin{equation}
\sigma(x,u)=c(x)+a(x)u+\alpha\!\left(h(x)\right).
\end{equation}

Definition~\ref{definition:SmoothPenaltyFunction} introduces a penalty that depends jointly on $h(x)$—measuring proximity to the unsafe set—and on $\sigma(x,u)$, which encodes the instantaneous CBF margin. The penalty becomes active only when \emph{both} $h(x)<\delta$ and $\sigma(x,u)<\mu$ hold. Consequently, the safety filter remains completely inactive when the state lies outside the sensing region ($h(x)\ge \delta$), ensuring that the nominal controller is not unnecessarily altered far from the unsafe set.

\begin{rem}
Although the perception gate does not appear explicitly in the penalty definition, its effect is implicitly recovered through the dependence on $h(x)$ in $\psi$.
\end{rem}

\begin{prop}\label{proposition:penalty-filter-solution}
Let $\mathcal{X}$ be the safe set defined in \eqref{eq:safe set}.  
The solution of the unconstrained problem \eqref{eq:penalty-problem} is
\begin{equation}
u^\star(x)
=
u_0(x)-
\frac{\psi(h(x),\sigma(x))\,\sigma(x)}
{\,1+\psi(h(x),\sigma(x))\,\|a(x)\|_{W^{-1}}\,}\;\nu(x),
\label{eq:penalty-problem-solution}
\end{equation}
where $\sigma(x)$ and $\nu(x)$ are given by \eqref{eq:sigma_1} and \eqref{eq:v x}, respectively.
\end{prop}

\begin{pf}
    See Appendix \ref{appendix:C}.
\end{pf}

Proposition~\ref{proposition:penalty-filter-solution} yields a fully closed-form expression for the smooth penalty-based safety filter. In contrast to the perception-limited CBF filter, this formulation contains no case distinctions: the controller is obtained from a smooth optimization criterion and inherits the regularity of the penalty function. The following theorem establishes the regularity and forward-invariance properties of the resulting feedback law.

\begin{thm}\label{thm:penalty-based}
    Consider the set $\mathcal{X}$ defined by \eqref{eq:safe set} and the dynamical system \eqref{eq:sys}. The feedback control law $u^*$ given by \eqref{eq:penalty-problem-solution} is locally Lipschitz-continuous, and if $f(x)$ and $g(x)$ are of class $\mathcal{C}^i$ and $\mathcal{C}^j$, respectively, then the controller $u^*$ is of class $\mathcal{C}^{\min\{k,l-1,p,r,i,j\}}$. Under the controller $u^*$, the dynamical system \eqref{eq:sys} admits a unique solution and the set $\mathcal{X}$ is forward invariant.
\end{thm}

\begin{pf}
See Appendix~\ref{appendix:D}.
\end{pf}

Theorem~\ref{thm:penalty-based} shows that the penalty-based controller preserves both smoothness and invariance guarantees. In contrast to the perception-limited filter based gated QP, which relies on piecewise switching that limits it to a locally Lipschitz continuous control law, the penalty-based formulation yields a controller that is differentiable up to the minimum smoothness of its constituent functions. This property makes the approach particularly suitable for applications requiring higher-order system models. 
Moreover, the forward invariance result shows that safety is preserved even as the smooth penalty function $\psi(h(x),\sigma(x))$ grows unbounded near constraint violation, ensuring that the safe set $\mathcal{X}$ remains invariant.  

\begin{rem}\label{remark:PsiChoice}
The choice of the function $\psi(h(x), \sigma(x))$ can be arbitrary as long as it satisfies Definition~\ref{definition:SmoothPenaltyFunction}. However, when $\psi$ grows unbounded as $h(x)$ and $\sigma(x)$ approach zero, numerical instabilities may arise despite the bounded nature of the resulting control law \eqref{eq:penalty-problem-solution}. To solve such issues, we introduce an alternative formulation of $\psi$ that retains the desired gating properties while improving numerical conditioning.
Specifically, we define
\begin{equation}
    \psi(h(x), \sigma(x)) = 
    \frac{\phi_\delta(h(x))\,\phi_\mu(\sigma(x))}{\|a(x)\|_{W^{-1}}\big(1 - \phi_\delta(h(x))\,\phi_\mu(\sigma(x))\big)},
\end{equation}
where
\begin{equation}
    \phi_\tau(z) :=
    \begin{cases}
        0,  & z \in [\tau, +\infty),\\
        \varphi(z), & z \in (0, \tau),\\
        1, & z \in (-\infty, 0],
    \end{cases}
\end{equation}
and $\phi_\tau(\cdot)$ is a smooth transition function parameterized by $\tau > 0$. The function $\varphi(\cdot) \in [0,1]$ ensures a sufficiently differentiable transition between $1$ and $0$ over $(0,\tau)$. For instance, using a cubic polynomial satisfying $\varphi(0)=1$, $\varphi'(0)=0$, $\varphi(\tau)=0$, and $\varphi'(\tau)=0$, one obtains $\varphi(z)=1-3(z/\tau)^2+2(z/\tau)^3$.

The term $\phi_\delta(h(x))$ provides smooth regularization with respect to the safety margin $h(x)$, whereas $\phi_\mu(\sigma(x))$ smooths the evaluation with respect to the constraint term $\sigma(x)$. 

With this formulation, the control law \eqref{eq:penalty-problem-solution} can be equivalently written as
\begin{equation}
    u^*(x) = u_0(x)-\phi_\delta(h(x))\phi_\mu(\sigma(x))\frac{\sigma(x)}{\|a(x)\|_{W^{-1}}}\nu(x),
\end{equation}
thereby preventing numerical divergence while maintaining smoothness and continuity across the safety set.
\end{rem}

The preceding analysis established that the penalty-based filter guarantees forward invariance of the safe set $\mathcal{X}$. 
However, forward invariance alone ensures that trajectories remain within $\mathcal{X}$ but does not imply any form of convergence toward the safe set. In practice, it is recommended when starting slightly away from the safe set (\emph{i.e.,} in the unsafe set) to be able to converge back to it. To formalize this notion, we adopt the framework of \emph{zeroing barrier function (ZBFs)} introduced in \cite{Ames2017}, which extends the concept of forward invariance to cover \emph{asymptotic set stability}. 
In our case, the smooth penalty-based filter induces barrier dynamics effects and possess the required conditions for asymptotic convergence, as shown next.

\begin{prop}\label{thm:asymptotic-stability-hobf}
Consider the control-affine system~\eqref{eq:sys} and let the safe set $\mathcal{X}$ be defined by~\eqref{eq:safe set}. Under the penalty-based controller~\eqref{eq:penalty-problem-solution}, the set $\mathcal{X}$ is asymptotically stable.
\end{prop}

\begin{pf}
To prove this property, one must show that $h(x)$ is a ZBF for the dynamical system \eqref{eq:sys}. We proved in Theorem~\ref{thm:penalty-based} the forward invariance of the set $\mathcal{X}$, hence $h(x)\ge 0$ for $x\in\mathcal{X}$. It also follows that $c(x)+a(x)u^*(x)+\alpha(h(x))\ge0$ on $x\in\mathcal{X}$ since the penalty-based controller \eqref{eq:penalty-problem-solution} is solution of the optimization problem \eqref{eq:penalty-problem}. Since $\dot{h}(x) = c(x)+a(x)u^*(x)$, we can easily say that $h(x)$ is a ZBF for the set~\eqref{eq:safe set} as defined in \cite[Definition~3]{Ames2017}. This satisfies the sufficient condition for asymptotic set stability stated in~\cite[Proposition~2]{Ames2017}. Consequently, the set $\mathcal{X}$ is asymptotically stable.
\end{pf}

Proposition~\ref{thm:asymptotic-stability-hobf} establishes that the penalty-based filter not only preserves forward invariance but also guarantees asymptotic convergence toward the safe set $\mathcal{X}$. This result is particularly relevant when the system starts slightly outside $\mathcal{X}$. In such situations, forward invariance alone is insufficient, as it only ensures that trajectories already inside $\mathcal{X}$ never leave it. The additional requirement is that trajectories outside the set evolve so that $h(x(t))$ increases toward nonnegative values as $t \to \infty$.

\begin{rem}[Extension to Multiple Unsafe Sets]\label{remark:multiple-penalties}

The proposed penalty-based framework can be readily extended to handle multiple sensed unsafe regions, each characterized by an individual barrier function $h_i:\R^n\to\R$, $i\in\{1,\dots,N\}$. 
Let $\psi_i(h_i(x),\sigma_i(x))$ denote the smooth penalty function associated with the $i$-th constraint, where $\sigma_i(x)=L_fh_i(x)+L_gh_i(x)u_0(x)+\alpha_i(h_i(x))$. 
The overall optimization problem is then defined as
\begin{equation}\label{eq:multi-penalty}
    \min_{u\in\R^m}\ \frac12\norm{u-u_0(x)}_W^2
    + \frac12\sum_{i=1}^{N} P_i(x,u),
\end{equation}
with
$
    P_i(x,u) := \psi_i\big(h_i(x),\sigma_i(x)\big)
    \big[\sigma_i(x,u)\big]^2.
$
Each penalty term $P_i$ activates smoothly only when the corresponding \emph{i}th unsafe set becomes perceptible. Hence, the filter automatically modulates the contribution of each constraint based on how far is the unsafe set. 
Importantly, the sum structure preserves convexity of the cost function and ensures that the overall control law remains smooth, as all terms are differentiable and bounded by construction. This property is difficult to retain when stacking CBF constraints for each $h_i$, as the resulting QP may lose smoothness or become infeasible when multiple constraints activate. In contrast, the penalty formulation ensures a globally defined and smooth control law.
\end{rem}

\section{Simulation Results}\label{sec:simulation}
We validate the proposed penalty-based safety filter on different systems. We first consider a first-order dynamics model to design a penalty filter, and then show that the same penalty filter can be accurately tracked by both fully actuated and underactuated second-order systems. We start by considering the first-order dynamics
\begin{equation}
    \dot{x} = v,
\end{equation}
which serves only to construct the safety-filtered velocity command that will be tracked by the higher-order systems in the subsequent examples. The nominal controller is a proportional law that drives the robot to the desired position $x_d$ and is given by
\begin{equation}
    u_0(x)=-k(x-x_d),
\end{equation}
where $k>0$ is a positive real. We design our controller to avoid collision with a circular obstacle centered at $c=(c_x,x_y)$ of a radius $r_c$. The safety function is defined as:
\begin{equation}
    h(x) = \|x-c\|-(r_c+\epsilon),
\end{equation}
where $\epsilon$ is a security distance. The class-$\mathcal{K}$ function is selected as $\alpha(h)=\alpha_0 h$, where $\alpha_0>0$ is a positive real. The perception gate is defined as a cubic Hermite window satisfying
\begin{equation}
\gamma(h) =
\begin{cases}
1, & h \le 0,\\[1pt]
1-3\Big(\tfrac{h}{\delta-\epsilon}\Big)^2 + 2\Big(\tfrac{h}{\delta-\epsilon}\Big)^3, & 0 < h < \delta-\epsilon,\\[1pt]
0, & h \ge \delta-\epsilon,
\end{cases}
\end{equation}
Based on what was discussed in Remark~\ref{remark:PsiChoice}, we adopt the alternative formulation to define the penalty function, given two positive reals $\delta$ and $\mu$. With these preliminaries, we can compute the required velocity $v=u^*(x)$, where $u^*(x)$ is given by \eqref{eq:penalty-problem-solution}. For higher-order dynamics, we modify the controller so that the robot's velocity tracks the penalty-based controller $v = u^*(x)$. We study two examples: a fully actuated robot and an underactuated planar drone.

\begin{figure*}[t]
    \centering
    \begin{subfigure}[t]{0.22\linewidth}
        \centering
        \includegraphics[width=\linewidth]{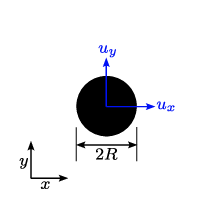}
        \caption{}
    \end{subfigure}%
    ~ 
    \begin{subfigure}[t]{0.22\linewidth}
        \centering
        \includegraphics[width=\linewidth]{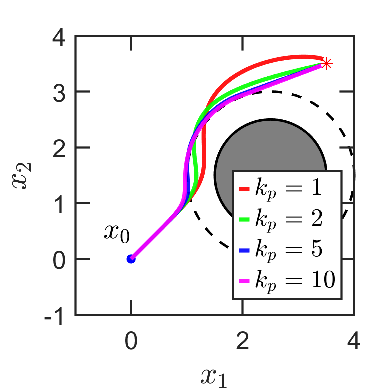}
        \caption{}
    \end{subfigure}
    ~ 
    \begin{subfigure}[t]{0.22\linewidth}
        \centering
        \includegraphics[width=\linewidth]{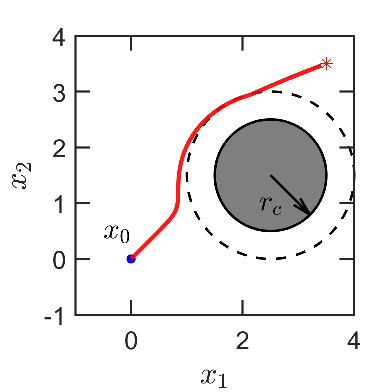}
        \caption{}
    \end{subfigure}
    \caption{Simulation results when applying the controller on a fully actuated robot. (a) A diagram of a fully actuated robot in 2D. (b) Resulting trajectories when tracking the computed penalty-based controller in the absence of the feedforward term. (c) Resulting trajectory when the feedforward term is applied, with the gain $k_p=1$.}
    \label{fig:tracking1}
\end{figure*}

\textit{1) Fully-Actuated Robot:} we consider the second-order dynamics
\begin{equation}
    \ddot{x}=u.
\end{equation}
To track the safety-filtered controller $u^*(x)$, we use 
\begin{equation}
    u = -k_p(\dot{x}-u^*(x))+\frac{\partial u^*(x)}{\partial x}\dot{x},
\end{equation}
where $k_p>0$ is a positive gain. This example highlights the main practical benefit of our approach: because $u^*(x)$ is continuously differentiable, in contrast to controllers issued from classical CBFs, the feedforward term $\frac{\partial u^*}{\partial x}\dot{x}$ can be computed explicitly, enabling accurate tracking with moderate gains. Figure~\ref{fig:tracking1} shows both cases, when the feedforward term is employed and when it is not. One can easily observe the improvement: when the feedforward term is included, the velocity tracks the penalty filter using a small gain $k_p$, whereas in its absence significantly larger gains are required to obtain comparable performance.

\begin{figure*}[t]
    \centering
    \begin{subfigure}[t]{0.22\linewidth}
        \centering
        \includegraphics[width=\linewidth]{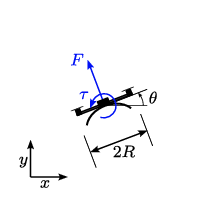}
        \caption{}
    \end{subfigure}%
    ~ 
    \begin{subfigure}[t]{0.22\linewidth}
        \centering
        \includegraphics[width=\linewidth]{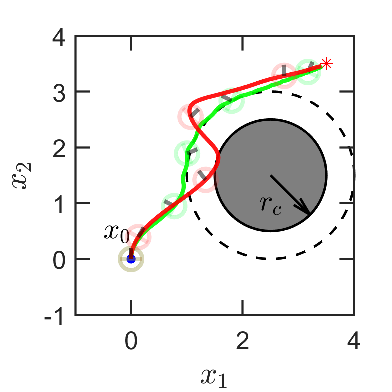}
        \caption{}
    \end{subfigure}
    ~ 
    \begin{subfigure}[t]{0.22\linewidth}
        \centering
        \includegraphics[width=\linewidth]{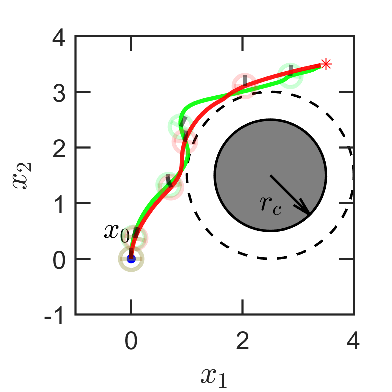}
        \caption{}
    \end{subfigure}
    \caption{Simulation results when applying the controller on a planar. (a) A diagram of a planar drone. (b) Resulting trajectories of the drone tracking the computed penalty-based controller when the feedforward term is not applied. The (green) trajectory corresponds to the gains $k_{\theta}=150$, $k_{\omega} = 10$ and $k_v = 10$, while the (red) trajectory corresponds to the gains $k_{\theta}=10$, $k_{\omega} = 2$ and $k_v = 1$. (c) Resulting trajectories with the feedforward term. The (green) trajectory corresponds to the gains $k_{\theta}=2$, $k_{\omega} = 2$ and $k_v = 1$, while the (red) trajectory corresponds to the gains $k_{\theta}=5$, $k_{\omega} = 5$ and $k_v = 1$.}
    \label{fig:tracking2}
\end{figure*}

\textit{2) Planar Drone:} for a planar drone with the dynamics
\begin{equation}
    \ddot{x}_1 = -\frac{F}{m}\sin{(\theta)},\quad
    \ddot{x}_1 = \frac{F}{m}\cos{(\theta)}-g_v,\quad
    \ddot{\theta} = \frac{\tau}{J},
\end{equation}
where $\theta\in\R$ is the pitch angle, $F\in\R$ is the thrust input, $\tau\in\R$ is the torque input, $g_v$ is the gravity, $m$ is the mass of the robot and $J$ is its moment of inertia. The control law is organized hierarchically. An outer-loop controller computes a desired acceleration $a_d$ that drives the velocity error to zero, while an inner-loop controller converts this desired acceleration into thrust and attitude commands. The outer-loop controller is expressed as 
\begin{equation}
    a_d(x) = -k_v(\dot{x}-u^*(x))+\frac{\partial u^*(x)}{\partial x}\dot{x},
\end{equation}
where $k_v>0$ is a positive gain. From 
$a_d$, the corresponding thrust and desired orientation are determined as
\begin{align}
    F &= m\|a_d+g_v e_y\|,\\
    \theta_d &= \arctan2(-a_{d,x},a_{d,y}+g_v),
\end{align}
where $e_y = [0,1]^\top$ is the vertical unit vector. The actual attitude is then driven to $\theta_d$ using a PD law
\begin{equation}
    \tau = -J(k_{\theta}(\theta-\theta_d)+k_{\omega}(\dot{\theta}-\dot{\theta}_{d})),
\end{equation}
where $k_{\theta},k_{\omega}>0$ are positive gains. Again, this example illustrates the advantages of having a smooth penalty filter $u^*(x)$. Tracking $u^*(x)$ requires not only the Jacobian $\frac{\partial u^*}{\partial x}$ but also the time derivative of the desired orientation $\dot{\theta}_d$, which depends on higher-order derivatives of $u^*(x)$ (the explicit expression of $\dot{\theta}_d$ is omitted here for clarity). As shown in Figure~\ref{fig:tracking2}, analytically computing both $\frac{\partial u^*}{\partial x}$ and $\dot{\theta}_d$ significantly improves tracking performance. In contrast, omitting these terms necessitates considerably higher gains $k_v$, $k_{\theta}$, and $k_{\omega}$ to obtain comparable behavior, which amplifies disturbances and degrades overall performance (as illustrated more clearly in the provided videos).

Videos of the simulations can be viewed from the links in the footnote\footnote{\url{https://youtu.be/My3b2mqq538}\\ \url{https://youtu.be/-NKosbsrdK4}}

\section{Conclusion}\label{sec:conclusion}
In this paper, we developed a smooth safety filtering framework for nonlinear control-affine systems operating under perception limitations. By introducing smooth perception gating, we generalized the classical CBF formulation to account for partial perception while maintaining forward invariance of the safe set. To overcome the nonsmoothness of the switching mechanism resulting from QP-based formulations, we further introduced a smooth penalty-based filter that blends the nominal and safe modes through smooth nominal gating function. This formulation preserves safety while achieving smooth closed-form controller. Future works will focus on rigorously characterizing the stability properties of the closed-loop dynamics under the proposed filters.

\bibliography{ifacconf}             
                                                   







\appendix

\section{Proof of Theorem~\ref{thm:gated-invariance}}\label{appendix:B}
\begin{pf}
We begin by recalling several preliminary facts that follow directly from the assumption on the regularity of all the elementary functions defined previously. The nominal controller $u_0(x)$ is smooth, hence it is locally Lipschitz-continuous, \emph{i.e.,} $\forall x_0\in\mathcal{X},\exists\rho>0,L_{u_0}>0$ such that $\forall x,y\in\mathcal{B}(x_0,\rho)$,
\begin{equation}
    \|u_0(y)-u_0(x)\|\le L_{u_0}\|y-x\|.
\end{equation}
The term $\|a(x)\|_{W^{-1}}$ admits a real $m\neq 0$ such that $m:=\inf_{x\in\mathcal{X}}\|a(x)\|_{W^{-1}}$. Since $a(x)$ and $c(x)$ are a composition of smooth and locally Lipschitz-continuous functions, then the functions $\sigma(x)$ and $\nu(x)$ are also locally Lipschitz-continuous. Hence, there exist two positive reals $L_\nu$ and $L_{\sigma}$, such that: $\|\sigma(y)-\sigma(x)\|\le L_{\sigma}\|y-x\|$ and $\|\nu(y)-\nu(x)\|\le L_{\nu}\|y-x\|$. Let $\sigma_{\max}:=\sup_{x\in\mathcal{X}}\sigma(x)$ and $\nu_{\max}:=\sup_{x\in\mathcal{X}}\nu(x)$.
To prove that the controller \eqref{eq:gated-qp-solution} is locally Lipschitz, we first define the set $\mathcal{S}$ such that:
\begin{gather}
    \mathcal{S}:=\{x\in\R^n|\gamma(h(x))\sigma(x)<0\}.
\end{gather}
Let $x\in\mathcal{X}$ and $y\in\mathcal{X}$ be two distinct points such that there exist a real $p\in[0,1]$ such that $x_p$ lies in $\mathcal{X}$ and defined as follows:
\begin{equation}
    x_p = px + (1-p)y.
\end{equation}
We distinguish between three cases:
Case 1: The points $x$ and $y$ are not in $\mathcal{S}$. Then, we have that:
    \begin{align*}
        \|u^*(y)-u^*(x)\|=\|u_0(y)-u_0(x)\|\le L_{u_0}\|y-x\|.
    \end{align*}
Case 2: Both points $x$ and $y$ are in $\mathcal{S}$. In this case one has:
    \begin{align*}
        &\|u^*(y)-u^*(x)\|\\
        &=\Big\|\big(u_0(y)-u_0(x)\big)-\Big(\tfrac{\sigma(y)}{\|a(y)\|_{W^{-1}}}\nu(y)-\tfrac{\sigma(x)}{\|a(x)\|_{W^{-1}}}\nu(x)\Big)\Big\|,\\
        &\le L_{u_0}\|y-x\|+\frac{1}{m}\|\sigma(y)\nu(y)-\sigma(x)\nu(x)\|,\\
        &\le L_{u_0}\|y-x\| + \tfrac{\|\sigma(y)\|}{m}\|\nu(y)-\nu(x)\|+\tfrac{\|\nu(x)\|}{m}\|\sigma(y)-\sigma(x)\|,\\
        &\le (L_{u_0}+\frac{1}{m}\big(\sigma_{\max}L_\nu+\nu_{\max}L_{\sigma})\big)\|y-x\|.
    \end{align*}
Case 3: $x$ lies in $S$ but $y$ does not. In this case, we define $p^*$ by:
\begin{equation}
    p^* := \min_{p \in [0,1],\, x_p \notin \mathcal{S}} p.
\end{equation}
When $p\in[0,p^*)$, then $x_p$ lies in $\mathcal{S}$ and as a result, we have that:
\begin{multline*}
    \|u^*(x_p)-u^*(x)\| \le (L_{u_0}+\frac{1}{m}\big(\sigma_{\max}L_\nu+\nu_{\max}L_{\sigma})\big)\|y-x\|.
\end{multline*}
While if $p\in[p^*,1]$, then $x_p$ is not in $\mathcal{S}$ and as a result, we have that:
\begin{align*}
    \|u^*(y)-u^*(x_p)\|\le L_{u_0}\|y-x\|.
\end{align*}
This leads to the following result:
\begin{align*}
    \|u^*(y)-u^*(x)\|\le L_{u^*}\|y-x\|,
\end{align*}
where $L_{u^*}=(2L_{u_0}+\frac{1}{m}\big(\sigma_{\max}L_\nu+\nu_{\max}L_{\sigma})\big)$.
Finally, we conclude that the control law \eqref{eq:gated-qp-solution} is locally Lipschitz continuous.

Next, we prove that the dynamical system \eqref{eq:sys} under the controller \eqref{eq:gated-qp-solution} admits a unique solution. We write the closed-loop system as:
\begin{equation}
    \dot{x} = f(x) + g(x)u^*,
\end{equation}
which right-hand side is locally Lipschitz-continuous on $\mathcal{X}$. We now show that any trajectory starting in $\mathcal{X}$ stays in it for all future time. We assume by contradiction that there exists $t_1>0$ with $h(x(t_1))<0$. We define
\begin{equation*}
    t_0:=\inf_{t\in[0,t_1]} h(x(t))<0.
\end{equation*}
At $t_0$ we have $h(x(t_0))=0$ and $h(x(t))\ge0$ for all $t\in[0,t_0]$. And since the function $h(x(t))$ decreases immediately for $t>t_0$, then $\dot{h}(x(t))<0$. From the other hand, the controller $u^*$ is solution of the gated QP and always satisfies the inequality
\begin{equation}\label{eq:tmp1}
\gamma\!\big(h(x)\big)\,\Big(c(x)+a(x)u^*(x)+\alpha\!\big(h(x)\big)\Big)\ \ge\ 0.
\end{equation}
At the boundary of $\mathcal{X}$, we have that $h(x(t_0)) = 0$, $\alpha(h(x(t_0)))=\alpha(0)=0$ and $\gamma(h(x(t_0)))=\gamma(0)=1$. Therefore, the inequality \eqref{eq:tmp1} is written at the boundary as
\begin{equation}
    c(x(t_0))+a(x(t_0))u^*(x(t_0))=\dot{h}(x(t_0))\ge 0.
\end{equation}
This contradicts the definition of $t_0$ as the time when $\dot{h}(x_0)$ is strictly negative. Therefore, there exists no such $t_1$ and $h(x(t))\ge 0$ for all $t\ge 0$. Thus, the trajectories will stay inside or at the boundary of the set $\mathcal{X}$. With this, we met the conditions of \cite[theorem 3.3]{khalil2002nonlinear} and as a result the closed-loop system admits a unique solution. Eventually, we showed that the set $\mathcal{X}$ is positively invariant under the proposed control law \eqref{eq:gated-qp-solution} given the above discussion.

\end{pf}

\section{Proof of Proposition~\ref{proposition:penalty-filter-solution}}\label{appendix:C}
\begin{pf}
    Considering the unconstrained optimization problem \eqref{eq:penalty-problem}, which is strictly convex and admits a unique solution. Let $\mathcal{L}(u)$ denote the cost function defined as follows:
    \begin{multline}
        \mathcal{L}(u)=\frac12\norm{u-u_0}_{W}^2
         +\ \frac12\,\psi(h,\sigma)\,\big(c+au+\alpha(h)\big)^2.
    \end{multline}
    To find the optimal control input, we compute the gradient of $\mathcal{L}(u)$ with respect to \( u \):
    \begin{multline}
        \frac{\partial\mathcal{L}}{\partial u}=W(u-u_0)
        +\ \psi(h,\sigma)\big(c+au+\alpha(h)\big)a^\top.
    \end{multline}
    We set $\frac{\partial\mathcal{L}}{\partial u}=0$ and rearrange the equation to obtain:
    \begin{multline}\label{eq:DL}
        (W+\psi(h,\sigma)a^\top a)u\\=Wu_0-\psi(h,\sigma)(c+\alpha(h))a^\top.
    \end{multline}
    To obtain the inverse of the term $(W+\psi(h,\sigma)a^\top a)$, we apply the Sherman-Morrison formula to obtain:
    \begin{multline}\label{eq:inverse-formula}
        (W+\psi(h,\sigma)a^\top a)^{-1}=W^{-1}\\-\frac{\psi(h,\sigma)W^{-1}a^\top aW^{-1}}{1+\psi(h,\sigma)aW^{-1}a^\top}.
    \end{multline}
    We plug \eqref{eq:inverse-formula} in \eqref{eq:DL} and solve for $u$ to obtain following optimal controller \eqref{eq:penalty-problem-solution}.

\end{pf}

\section{Proof of Theorem~\ref{thm:penalty-based}}\label{appendix:D}
\begin{pf}
    The smoothness of the controller \eqref{eq:penalty-problem-solution} depends directly on the classes of the elementary functions that defines it. For instance, one has that:
    \begin{align}
        \mathcal{C}(a)&=\min(\mathcal{C}(h)-1,\mathcal{C}(g))=\min(l-1,j),\\
        \mathcal{C}(c)&=\min(\mathcal{C}(h)-1,\mathcal{C}(f))=\min(l-1,i).
    \end{align}
    Since the denominator $(1+\psi(h(x),\sigma(x))\|a(x)\|_{W^{-1}})\neq 0$, then the class of the controller $u^*$ is given as:
    \begin{align*}
        \mathcal{C}(u^*)&=\min(\mathcal{C}(u_0),\mathcal{C}(\psi(h,\sigma)),\mathcal{C}(s),\mathcal{C}(\nu)),\\
        &=\min(\mathcal{C}(u_0),\mathcal{C}(\psi),\mathcal{C}(h),\mathcal{C}(a),\mathcal{C}(c),\mathcal{C}(\alpha)),\\
        &=\min(k,l-1,p,r,i,j).
    \end{align*}
    We prove the invariance of the set $\mathcal{X}$ by first computing the derivation of $h(x)$ by substituting the controller \eqref{eq:penalty-problem-solution} in \eqref{eq:hdot}:
    \begin{multline}
        \dot{h}(x)=c(x)+a(x)u_0(x)\\-\dfrac{\psi(h(x),\sigma(x))\sigma(x)}{1+\psi(h(x),\sigma(x))\|a(x)\|_{W^{-1}}}\,\|a(x)\|_{W^{-1}}.
    \end{multline}
    We recall that $\sigma(x)=c(x)+a(x)u_0(x)+\alpha(h(x))$ and we rewrite the $\dot{h}$-equation as follows:
\begin{multline}
        \dot{h}(x)=\sigma(x)-\alpha(h(x))\\-\dfrac{\psi(h(x),\sigma(x))\sigma(x)}{1+\psi(h(x),\sigma(x))\|a(x)\|_{W^{-1}}}\,\|a(x)\|_{W^{-1}}.
\end{multline}
On the boundary of $\mathcal{X}$, \emph{i.e.,} $h(x) = 0$, we have that $\alpha(h(x))=0$. Hence,
\begin{equation}
    \dot{h}(x)=\dfrac{\sigma(x)}{1+\psi(h(x),\sigma(x))\|a(x)\|_{W^{-1}}}.
\end{equation}
If $\sigma(x)\ge0$, then we clearly have $\dot{h}(x)\ge 0$. While if $\sigma(x)<0$, then $\psi(0,\sigma(x))\to +\infty$, therefore $\dot{h}(x)\to0^+$. Hence, $\dot{h}(x)\ge 0$ in all cases and for all $x$ in the boundary of $\mathcal{X}$. Finally, by Nagumo's theorem, all the trajectories starting in $\mathcal{X}$ remain in it for all $t\ge0$. 
\end{pf}
\end{document}